\documentclass[final,5p,times,twocolumn,draft]{elsarticle}

\usepackage{amssymb}
\usepackage{amsmath}
\usepackage{bm}
\usepackage{mathrsfs}
\usepackage{textcomp}
 \usepackage{placeins}
\usepackage{comment}
\usepackage{natbib}
\biboptions{sort&compress}
\usepackage{setspace}

\usepackage[dvipsnames]{xcolor}

\renewcommand{\eqref}[1]{\textup{{\normalfont Eq.~(\ref{#1}}\normalfont)}}

\newcommand{\hb}[1]{\hat{\bm{#1}}}

\usepackage{ulem}
\usepackage{xcolor}

\DeclareRobustCommand{\Erase}{\bgroup\markoverwith{\textcolor{red}{\rule[.5ex]{2pt}{0.4pt}}}\ULon}

\journal{Extreme Mechanics Letter}

\begin{document}

\begin{frontmatter}

\cortext[cor1]{Corresponding author. E-mail address: sano@mech.keio.ac.jp (T. G. Sano).}

\title{Twist deformation of physical trefoil knots}

\author[a]{Taiki Goto}
\author[a]{Shunsuke Nomura}
\author[a,b]{Tomohiko G. Sano{$^{*}$}}

\address[a]{School of Integrated Design Engineering, Graduate School of Science and Technology, Keio University,3-14-1 Hiyoshi, Yokohama, 2238522, Japan}
\address[b]{Department of Mechanical Engineering, Keio University,3-14-1 Hiyoshi, Yokohama, 2238522, Japan}


\begin{abstract}
Knots across various length scales, from micro to macro-scales, such as polymers, DNA, shoelaces, and surgery, serving their unique mechanical properties. 
The shape of ideal knots has been extensively studied in the context of knot theory, while that of physical knots has been discussed very recently.
The complex interplay of elasticity and geometry, such as bending, twisting, and contact, needs to be disentangled to predict their deformation.
Still, the unified understanding of the deformation of physical knots is insufficient. 
Here, we focus on the trefoil knot, a closed knot with a nontrivial topology, and study the relationship between the shapes of the trefoil knot and applied physical twists, combining experiments and simulations. 
As we twist the elastomeric rod, the knot becomes either tightened or loosened, preserving the original three-fold symmetry, and then buckles and exhibits symmetry breaking at critical angles. The curvature profiles computed through the X-ray ($\mu$CT) tomography analysis also exhibit similar symmetry breaking. The transition would be triggered by the mechanical instability, where the imposed twist energy is converted into the bending energy. The phase transition observed here is analogous to the classical buckling phenomena of elastic rings known as the Michell instability.
We find that the twist buckling instability of the trefoil knot results from the interplay of bending, twisting, and contact properties of the rod. In other words, the buckling of the knot is predictable based on the elasticity and geometry of rods, which would be useful in avoiding or even utilizing their buckling in practical engineering applications such as surgery and the shipping industry.
\end{abstract}


\end{frontmatter}


\section{Introduction}
\label{Introduction}

Knots are formed when slender filament-like objects intertwine \cite{Ashley1944} and have a wide range of applications, from everyday uses such as shoelaces \cite{Daily-Diamond2017}, fishing \cite{hunter2021}, and rope-work \cite{dedekam2017} to natural science and engineering applications, for instance, surgical suturing techniques \cite{BURKHART2000, Bryan2014, johanns2023}, polymer science \cite{Saitta1999}, and DNA structures \cite{Ishihara2011, Tobias2000}. 
Characteristic mechanical properties of knots, \textit{e.g.,} in reducing tension along filaments or securing them in place, are utilized across various length scales, from the macro-scales to the micro-scales. 
Knot theory has a long history of research in mathematics \cite{gordon2006, crowell2012, kauffman2021}, while mechanics of physical knots have been studied in the last few decades.
Physical knots involve contact during tying, leading to friction and cross-sectional deformation, which can significantly influence their mechanical behavior. Consequently, much of the theoretical research thus far has focused on loose knots where cross-sectional deformation due to contact can be neglected \cite{Jawed2015, Audoly2007, Clauvelin2009}. 
Jawed \textit{et al.} conducted precise model experiments on loosely tied overhand knots, which are open knots tied by forming a loop and passing the end through it. They demonstrated excellent agreement between the experimental results and the analytical model derived from the nonlinear theory of slender elastic rods \cite{Jawed2015}. 
On the other hand, research on tightly tied knots, where contact cannot be neglected, has also advanced in the recent few years \cite{Baek2020, Sano2022, Grandgeorge2021, Grandgeorge2021cap, johanns2023, johanns2024}. 
Baek \textit{et al.} investigate the mechanical properties of tightly tied figure-eight and stopper knots using finite element analysis and experiments \cite{Baek2020}. Sano \textit{et al.} successfully quantify the geometric characteristics of complex knots through experiments and analysis on clove hitch knots \cite{Sano2022}.

Among the various types of knots, we focus on the trefoil knot, which is one of the simplest non-trivial knots and has been extensively studied within the framework of knot theory. The trefoil knot is a closed knot formed by gluing the two ends of an overhand knot. 
Applications of the trefoil knot are found in such as DNA \cite{Du1994, Weber2013}, polymer science \cite{Dietrich1989, noel2010}, and cyclotron orbits \cite{zhang2019}. 
Johanns \textit{et al.} perform a compare-and-contrast investigation between physical and ideal trefoil knots by combining a detailed characterization with X-ray tomography, knot theory, and three-dimensional finite element simulations. 
Their findings imply that by evaluating the shape, self-contact areas, curvature, and cross-sectional deformation, knots may weaken the strength of the filament itself \cite{Johanns2021}.

In general, when rod-like structures are twisted, bending and twist interact in a complex manner \cite{olson2013, Romero2021}. While the general relationship between bending and twist in non-trivial physical knots has not been fully elucidated, the buckling instability of the trivial knot, \textit{i.e.,} ring is known as the Michell instability \cite{michell1889,Goriely2006}. 
Shape a naturally straight rod into a ring of radius $R(= \kappa^{-1})$ and then twist it.
When the twist $\tau$ is sufficiently small, the ring maintains a planar shape, and the rod is uniformly twisted along the centerline. 
When the twist $\tau$ exceeds the critical twist $\tau^*$:
\begin{equation}
    \tau^* = \sqrt{3}\frac{B}{C}\kappa,
    \label{eq:Michell}
\end{equation}
the ring buckles into the figure-of-eight configuration.
The critical twist angle, $\phi^*$, calculated from the critical twist as $\phi^* = 2\pi \kappa^{-1} \tau^*$, depends solely on the bending $B$ and twist rigidity $C$ of the rod. For example, the critical angle of the isotropic incompressible elastic material (Poisson's ratio $\nu=0.5$) is given by $\phi^*\simeq5.20\pi$ with the aid of $B/C = 2/(1+\nu)$. Goriely \textit{et al.} reformulate the Michell instability and generalize the framework into the case with the intrinsic curvatures \cite{Goriely2006}.
Thus, it is evident that bending and twisting are inseparable in rod-like structures.

\begin{figure}[h!]
        \centering
        \includegraphics[width=0.48\textwidth]{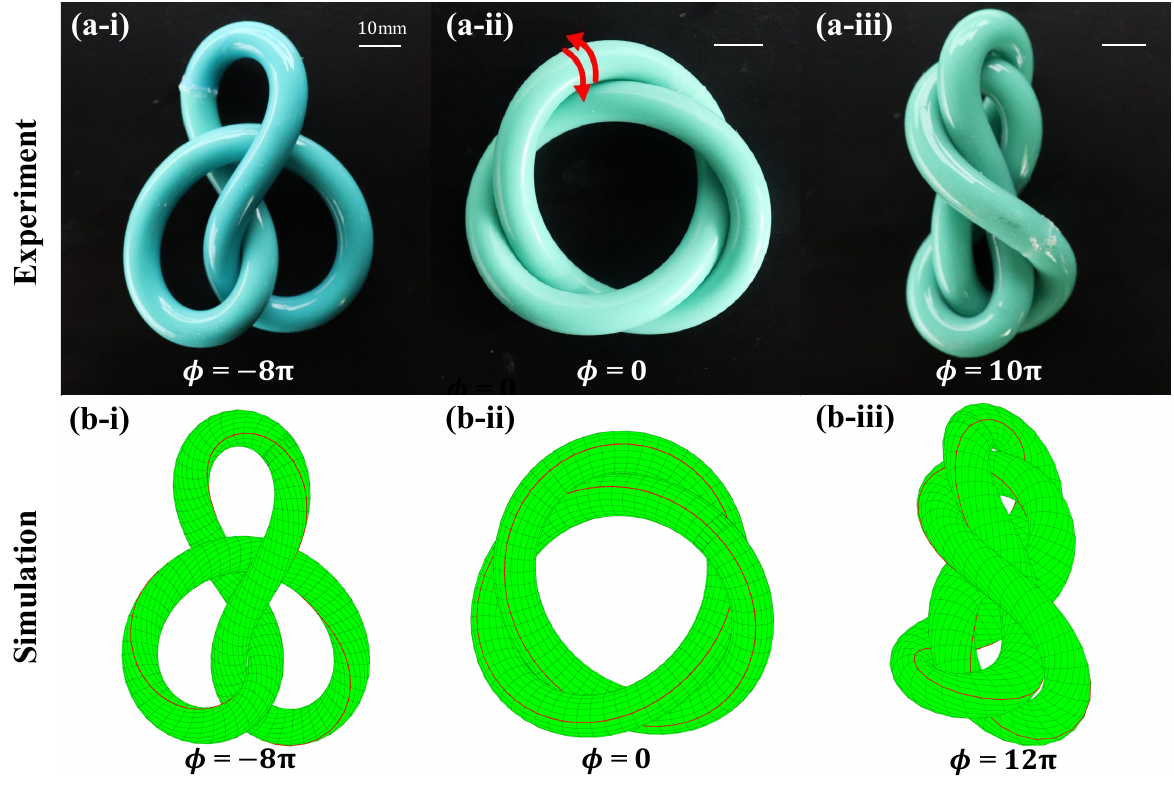}
        \caption{Twist deformation of elastic trefoil knots in experiment and simulation. 
        (a)Snapshots obtained from the experiments for a trefoil knot with $(D, L) = (8.2, 292)$~mm tied by imposing relative twist angle (a-i)$\phi=-8\pi$, (a-ii)$\phi=0$, (a-iii)$\phi=10\pi$.
        (b) Reconstructed configurations obtained in the discrete simulations with the twist angle (b-i)$\phi=-8\pi$, (b-ii)$\phi=0$, (b-iii)$\phi=12\pi$.
        The red curves drawn on the rod surface in the simulation represent the eccentric centerline parallel to the centerline, which allows visualization of the applied twists.
        }
        \label{Fig1}
\end{figure}

In this article, we clarify the relationship between the shapes of trefoil knots and applied physical twists, combining experiments and simulations. We investigate the morphological changes in the trefoil knot when the twist is applied. 
The trefoil knot is either loosened or tightened upon twist directions and then buckles above each critical angle.
We observe how the knot undergoes distinct buckling in each direction, both leading to symmetry breaking. 
To quantitatively evaluate the buckling phenomenon, we perform $\mu$CT analysis, and the characteristics of the buckling phenomenon are expressed with the topological numbers in space curves: writhing, twist and linking numbers.
We further estimate the critical angles for the transitions by applying the critical angle of the Michell instability; Eq.~(\ref{eq:Michell}), excellent agreement with experiments and simulations.

\section{Experimental protocols}
\label{sec:rod_fab}

\subsection{Fabrication of the rods for $\mu$CT}\label{sec:fab_uCT}
To visualize the three-dimensional geometry of a rod using $\mu$CT scanner (XDimensus 300, Shimadzu, Japan), we fabricate an elastic rod with composite materials following the methodologies outlined in Refs.~\cite{Grandgeorge2021,Sano2022}. 
The following is an overview of the rod fabrication procedure:
First, silicone elastomer (Elite Double 22, Zermack, Italy) is poured into an iron pipe with an inner diameter of 8 mm, into which a nylon fishing line (diameter: $0.9$ mm) is threaded at two positions along the centerline of the rod and an eccentric centerline parallel to it. After curing, the silicone elastomer is removed from the iron pipe, and the nylon fishing line is extracted to create a mold. This process produces a bulk elastic rod of 8~mm diameter, total length $L = 450$ mm, and the mass density, $\rho = 1.19$ g/cm$^3$, and the Young's modulus, $E = 0.474$ MPa with two thin holes, each with a diameter of $0.9$ mm.
Next, another silicone elastomer (Solaris, SMOOTH-ON, USA) is injected into the holes in a fiber-like hole using a syringe with an outer diameter of $1$ mm. Calcium carbonate (CaCO$_3$) is mixed into the silicone elastomer to increase its density (as $1.30$ g/cm$^3$), allowing it to be distinguished with $\mu$CT (See Sec.~\ref{sec:CL}).
Finally, the rod surface is dip-coated with the elastomer (Solaris, SMOOTH-ON, USA) by pulling the rod upward by $100$ mm/s using the force testing machine (EZ-LX, Shimadzu, Japan) from the liquid silicone bath. The cured coating layer has approximately 0.2 mm thickness, and the rod diameter, $D$, becomes $D=8.2$~mm. Given that the inner two fibers and the coating layer are sufficiently thin, they do not affect the mechanical performance of the composite rod~\cite{Grandgeorge2021}.
In the experiments conducted in this study, in addition to the composite elastic rods with a diameter of approximately $8.2$ mm fabricated using the above method, homogeneous silicone elastic rods (without any holes or coating layer) with a diameter of $4$ mm are also used (The details are provided in Sec. \ref{sec:geometry}).

\subsection{Tying the trefoil knot in the experiment}\label{sec:tying}
A trefoil knot can be formed by joining the two ends of an overhand knot. In this study, we introduce a twist into the trefoil knot by applying relative rotation of angle $2n\pi (n\in\mathbb{Z})$ to the cross-sections at rod-extremities before joining them. 
To realize an integer rotation and perfectly align the ends of the eccentric centerline, a short nylon fishing line (diameter: $0.9$ mm, length $\simeq 10$ mm) is embedded at the joint. 
The cross-sections are rotated as indicated by the red arrow at $\phi=0$ in Fig. \ref{Fig1}(a-ii), and then glued (Aron Alpha, Toagosei, Japan). 
The twisted trefoil knots for given $\phi$ are systematically fabricated through above procedure.
Finally, to account for potential asymmetry caused by gravity or friction, the fabricated knots are placed in an ultrasonic bath (Branson Ultrasonic Bath Bransonic CPX3800H-J, Yamato Scientific, Japan) for 5 minutes. This process is conducted to experimentally achieve a state of mechanical equilibrium, corresponding to the energetically most stable configuration as Ref.~\cite{Johanns2021}.

\subsection{Extraction of the Cosserat frame from the $\mu$CT data}\label{sec:CL}
After reconstructing the three-dimensional image of the trefoil knot from the cross-sectional data obtained using $\mu$CT, the coordinates of the centerline, $\bm{r}(s)$, and those of the eccentric centerline fibers, $\bm{r}_{\rm ex}(s)$, are binarized with the aid of the density difference between the elastomer with and without the CaCO$_3$ as described in Sec.~\ref{sec:fab_uCT}. 
The volumetric data of the centerline and the eccentric fibers are further skeletonized (reducing an object to a 1-pixel width curve) to approximate them by space curves.
The centerline coordinates $\bm{r}(s)$ are now extracted from the skeletonized centerline-fiber volume as a function of centerline arc length $s$. 
The Cosserat frame along the centerline is introduced for each arc length, $s$, as the local orthonormal basis, $(\hat {\bm{d}}_1, \hat {\bm{d}}_2, \hat {\bm{d}}_3)$ to calculate the curvature and twist of the rod~\cite{Audoly2010}. The experimental Cosserat frame on each centerline position, $\bm{r}(s)$, is constructed from the $\mu$CT data as follows~\cite{Sano2022}. The tangent vector of the centerline, $\bm{r}'(s) \simeq \{\bm{r}(s+ds) - \bm{r}(s)\}/ds$, is identified as $\hat{\bm{d}}_3 = \bm{r}'$ with the discretization length $ds\simeq0.07$mm. 
The vector $\hat {\bm{d}}_1(s)$ is defined as the direction of the intersection between the plane orthogonal to the centerline tangent, $\hat {\bm{d}}_3$, and the eccentric centerline, while the rest vector, $\hat {\bm{d}}_2(s)$, is computed as $\hat {\bm{d}}_2 \equiv \hat {\bm{d}}_3\times\hat {\bm{d}}_1$.
This protocol yields an orthonormal coordinate system (Cosserat frame) with $\hat {\bm{d}}_a(s)$($a=1,2,3$) as its basis vectors at each discrete point \cite{Audoly2010}.
The Cosserat frame evolves along the centerline with kinematic relation, $\hat{\bm{d}}_a ' = \bm{\Omega}\times\hat{\bm{d}}_a$, where the Darboux vector, $\bm{\Omega} = \kappa_1\hat{\bm{d}}_1 + \kappa_2\hat{\bm{d}}_2 + \tau\hat{\bm{d}}_3$ has the rotation rate of the Cosserat frame around $\hat{\bm{d}}_a$ as the $a$-th component of $\bm{\Omega}~(= \bm{\Omega}\cdot\hat{\bm{d}}_a)$.
The twist $\tau$ and the curvatures $\kappa_1$ and $\kappa_2$ of the elastic rod can be calculated from the kinematic relation as $\hat{\bm{d}}_a'\cdot\hat{\bm{d}}_b~(a, b = 1,2,3)$, i.e.,

\begin{eqnarray}
    {\kappa}_1(s) &=& \hat {\bm{d}}_2'(s)\cdot \hat {\bm{d}}_3(s)\label{eq:kappa1}\\
    {\kappa}_2(s) &=& \hat {\bm{d}}_3'(s)\cdot \hat {\bm{d}}_1(s)\label{eq:kappa2}\\
    \tau(s) &=& \hat {\bm{d}}_1'(s) \cdot \hat {\bm{d}}_2(s)\label{eq:tau}.
\end{eqnarray}
Given that the undeformed rod has a circular cross-section, the choice of $\hat{\bm{d}}_1$ is arbitrary. Hence, we introduce the magnitude of curvature, $\kappa(s)$ as Ref.~\cite{Sano2022}:
\begin{eqnarray}
    {\kappa}(s) = \sqrt{{{\kappa_1}}^2 + {{\kappa_2}}^2}\label{eq:kappa}.
\end{eqnarray}
The components of the Darboux vector, $\kappa_1, \kappa_2, \tau$, are computed by discretizing Eqs.~(\ref{eq:kappa1})-(\ref{eq:tau}) with the infinitesimal length $ds \simeq 0.07$mm. 



\section{Discrete simulations}
\label{sec:method_sim}

We perform the discrete simulation for twisting the trefoil knot. Our simulation is centerline-based (one-dimensional-like), neglecting the cross-sectional deformation~\cite{Chirico:1994hm}. The rod centerline, $\bm{r(s)}$, is discretized into a set of connected beads via stiff linear springs (The number of beads is set to be $N = 121$ throughout). The Cosserat frame, $\hb{d}_a~(a = 1,2,3)$, is attached to each bead, and the bending and twisting rigidity is taken into account by calculating the relative rotation of the adjacent Cosserat frames (through their Euler rotation). We consider the normal contact force to penalize self-crossings and to ensure the knotted configurations, while we neglect the sliding friction forces. The elastic energy functional of a naturally straight rod is discretized onto the variables of bead configurations, and its functional variation gives each internal force and torque. The coordinates of the beads and the torsional angle parameter are updated via Newton's equation of motion. The appropriate numerical dissipation (linear viscous damping) is also implemented to realize the mechanical equilibrium configurations within a reasonable computational time. The general protocols are detailed in \textit{e.g.,} Ref.~\cite{Sano2022sim}. Note that our simulation has no adjustable parameters except the damping parameters. 

The configuration of the twisted trefoil knot in the simulation is calculated as follows. The straight rod is bent into the three-fold symmetric curvature and twist by applying the modulation of wavelength $\simeq L/3$ for $\kappa_1$ and $\tau$.  
The two beads located at the rod extremities ($s = 0, L$) are then translated, sharing an identical location (to close the knot). Their Cosserat frames are further rotated around the centerline tangent, $\hb{d}_3$, to realize the zero-twist configuration ($\phi = 0$). By applying the Gaussian white noise (whose variance is of the order of the typical bending force)  and relaxing the rod for a sufficiently long time, we obtain the twistless trefoil knot ($\phi = 0$) in the discrete simulation. The twisted trefoil knot is realized by gradually imposing the relative twist angle, $\phi\ne0$, at the cross-section $s = 0, L$, together with the Gaussian white noise and the relaxation step as the twistless case.

\section{Shape transition of the twisted trefoil knot}
\label{sec:profile}

\subsection{Twist deformation of the trefoil knot}
We observe the morphological changes of the trefoil knot under twist angles $\phi\ne0$ both in experiments and simulations.
When the trefoil knot is twisted in one direction ($2\pi$,$4\pi$,…), the entire knot becomes tightly constrained and shrunk along the centerline, transitioning into a folded configuration at the critical angle, $\phi^*_+$(Fig.~\ref{Fig1}(a,b-ii)$\rightarrow$(a,b-iii)). 
Conversely, the knot loosens and ``opens" when twisted in the opposite direction ($-2\pi$,$-4\pi$,…), transitioning into a locally buckled configuration at the critical angle, $\phi^*_-$, as Fig.~\ref{Fig1}(a,b-ii)$\rightarrow$(a,b-i). 
We observe qualitatively similar phase transitions both in experiments and simulations (Fig.~\ref{Fig1}).
Based on these observations, we define the direction in which the knot tends to loosen (open) and tighten (close) as $\phi<0$ and $\phi>0$, respectively.

We observe the phase transition for $\phi>0$ at $\phi^*_+ \simeq 10\pi$ and $12\pi$ in the experiment and simulation of $d = 8$~mm, respectively.
In the range of $\phi<0$, the experiment and the simulation exhibit the transition at $\phi^*_- \simeq -8\pi$.
While the post-buckled configurations in experiments and simulations are qualitatively similar (a more detailed comparison will be discussed later), we observe a slight difference in critical angles for the phase transitions of experiments and simulations. We believe that the discrepancy originates from the simplified modeling in simulations, neglecting shear deformation. Indeed, the quantitative agreement is slightly improved for slender rods as $d = 4$~mm. See Sec.\ref{sec:geometry} for a more quantitative comparison.
From these observations, it is evident that when a twist is applied to the cross-section of the rod forming the trefoil knot, morphological transitions occur at critical angles. This finding suggests the existence of twist instability in the vicinity of these angles.

\subsection{Curvature and Twist profile of the twisted trefoil knot}

\begin{figure}[h!]
    \centering
    \includegraphics[width=0.48\textwidth]{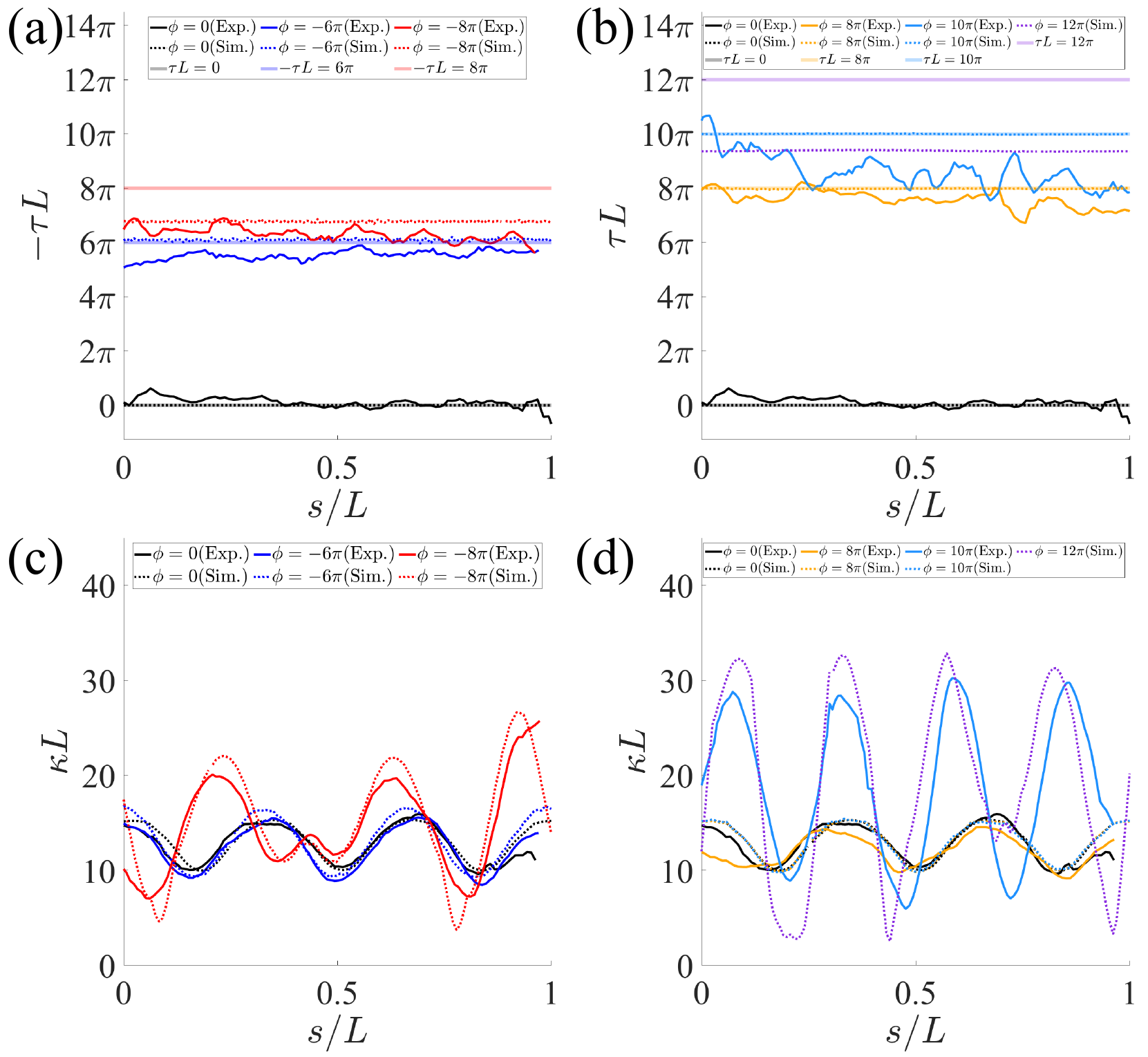}
        \caption{Profiles of (a)(b) the normalized twist, $\tau L$, and (c)(d) the normalized curvature,$\kappa L$, as functions of normalized arc-length, $s/L$, computed through the $\mu$CT analysis in the experiments (solid lines) and the discrete simulations (dotted lines). 
        We summarize the results for $\phi<0$ and $\phi>0$ in (a)(c) and (b)(d), respectively. 
        The shaded lines represent the corresponding applied twist angles $\phi$.
        }
        \label{Fig2}
\end{figure}

Utilizing the $\mu$CT experimental data (Sec.~\ref{sec:CL}) and simulation results, we analyze the three-dimensional geometry of the trefoil knot, curvature $\kappa$, and twist, $\tau$, in pre- and post-buckled configurations (Eqs.~(\ref{eq:tau}) and (\ref{eq:kappa})). We find that $\kappa$ and $\tau$ (normalized by the total length, $L$) profiles as functions of the normalized arclength, $s/L$, characterize the morphological transitions in the previous section. See Figs.\ref{Fig2}(a)(b) and (c)(d) for twist and curvature profiles, respectively.

We find that the twist is nearly constant along the rod centerline as $\tau(s) \simeq \tau_0 (= \rm{Const.})$, as shown in Figs.~\ref{Fig2}(a) and (b), irrespective of the buckling instability. The uniform profile of the twist can be understood from the general property of the centerline-based equilibrium balance equations (the Kirchhoff rod equations). In general, the twist of the rod with a circular cross-section is one of the integrals of the Kirchhoff rod equations in the absence of distributed torques~\cite{Audoly2010}. The twist in our configuration would be determined from the boundary condition (applied twist) and geometry of the centerline (torsion)~\cite{Coleman2004}.
Fig.~\ref{Fig2}(a)--(d) shows that the trefoil knot with no twist angle ($\phi=0$, black solid line) maintains a constant twist of zero, indicating that no unnecessary twist is introduced. The constant twist of the rod, $\tau_0$, varies along with the twist angle, $\phi$, until the buckling transition $\phi^*_-<\phi<\phi^*_+$, where the knot forms almost planar configuration maintaining the original three-fold symmetry as $\phi = 0$. Given that the knot remains almost planar, we expect that the constant twist and $\phi$ are related as $\tau_0 = \phi/L$ (shown as the shaded lines in Fig.~\ref{Fig2}), which are in excellent agreement with experiments and simulations except for the vicinity of the transition points, $\phi\simeq\phi^*_{\pm}$. When the magnitude of the twist angle exceeds the critical value, that of the constant twist, $|\tau_0|$, decreases abruptly, reflecting the out-of-plane buckling transition, where the twist elastic energy is converted into the bending one.

In the absence of the applied twist as $\phi=0$, the curvature profile has a three-fold symmetry consistent with the previous work~\cite{Johanns2021}. The identical symmetry is observed before the buckling, $\phi^*_-<\phi<\phi^*_+$ as expected. The curvature profile breaks its original symmetry above the transition points, while the broken symmetries are different for $\phi<0$ and $\phi > 0.$ When the trefoil knot is twisted to tighten it ($\phi>0$), the overall trefoil knot is folded into ``half" (See Fig.~\ref{Fig1}(a,b-iii)).  In this post-buckled configuration, the curvature profile has a nearly four-fold symmetry (appearing as four peaks), as shown in Fig.~\ref{Fig2}(d). On the other hand, when twisted to loosen the knot, the buckling is localized, forming a plectoneme-like structure (See Fig.~\ref{Fig1}(a,b-i)). The curvature profile for the buckled knot in $\phi<0$ forms a symmetric (two-fold symmetric) profile shown as Fig.~\ref{Fig2}(c). We have found that the applied twist triggers the shape transition in the trefoil knot, where the symmetry of the curvature profile changes from three-fold into four- or two-fold symmetric configurations.

\section{Writhing, Twist, Linking numbers of the twisted trefoil knot}
\label{sec:geometry}

We have shown that the trefoil knot undergoes shape transition upon twist. To clarify the geometric aspect of the transition in more detail, we introduce the geometric numbers known as writhing, $Wr$, twist, $Tw$, and linking, $Lk$ numbers~\cite{White1969,adams2004knot}. 
These numbers are often used in characterizing the entanglement of the DNA supercoilings and are computed from the configuration of two strands~\cite{Chirico:1994hm}. 
To compute the geometric numbers and characterize the shape transition, we quantify the topology of the twisted trefoil knot by considering the configuration of two curves in the rod: the rod centerline, $\bm{r}(s)$, and the eccentric centerline, $r_{\rm ex}(s)$. 
The writhing number describes the crossing of the rod centerline, $\bm{r}(s)$, defined in the continuum description as
\begin{eqnarray}
    \hspace{-0.1cm}
    Wr = \frac{1}{4\pi}\int ds'\int ds \left(\hb{d}_3(s)\times\hb{d}_3(s')\right)\cdot\frac{\bm{r}(s') - \bm{r}(s)}{|\bm{r}(s') - \bm{r}(s)|^3},\label{eq:Wr}
\end{eqnarray}
where the rod centerline $\bm{r}(s)$ is a non-self-intersecting curve in space with the arclength $s$. 
Note that only the configuration of the rod centerline is necessary to compute the writhing number. The twist number, $Tw$, is the total integral of the twist, 
\begin{eqnarray}
    Tw &=& \frac{1}{2\pi}\int ds ~\hb{d}_3(s)\cdot\left(\hb{d}_1(s)\times\hb{d}_1'(s)\right)\nonumber\\
    &=& \frac{1}{2\pi}\int ds~\tau(s)\label{eq:Tw},
\end{eqnarray}
where the number of turns of $\hb{d}_1$ around the tangent, $\hb{d}_3$, is counted. 
The (Gauss) linking number, $Lk$, counts the crossing of the centerline and the eccentric centerline, defined by the double integral as
\begin{eqnarray}
    \hspace{-0.5cm}
    Lk = \frac{1}{4\pi}\int ds'\int ds \left(\hb{d}_3(s)\times\hb{d}_3(s')\right)\cdot\frac{\bm{r}_{\rm ex}(s') - \bm{r}(s)}{|\bm{r}_{\rm ex}(s') - \bm{r}(s)|^3}\label{eq:Lk}.
\end{eqnarray}
It should be stressed that the linking number, $Lk$, is an invariant for closed space curves and is related to the writhing, $Wr$, and twist, $Tw$, as
\begin{eqnarray}
    Lk = Wr + Tw,\label{eq:lkwrtw}
\end{eqnarray}
known as the Calugareanu-White-Fuller theorem~(See Ref.~\cite{Dennis2005} and references therein). 

We compute the writhing, $Wr$, twist, $Tw$, and linking, $Lk$ numbers in our experimental and simulation results below (by discretizing their definitions Eqs.~(\ref{eq:Wr})-(\ref{eq:Lk})). To demonstrate their geometric meaning, we first consider the twistless trefoil knot ($\phi = 0$). By counting the crossing of the centerline, we get $Wr = 3$. Given that no twist is applied to the trefoil knot, the twist number is zero, $Tw = 0$. Hence, the linking number is calculated as $Lk = 3$. When the twist is applied $\phi\ne0$, the eccentric centerline is no longer a closed curve, and the linking number, $Lk$, is no longer an integer for general $\phi$. We expect that the linking number is written as
\begin{eqnarray}
    Lk = 3 + \frac{\phi}{2\pi},\label{eq:lkphi}
\end{eqnarray}
whereas the writhing, $Wr$, and twist, $Tw$, numbers would change, still satisfying Eq.~(\ref{eq:lkwrtw}). In other words, the writhing, $Wr$, and twist, $Tw$, numbers characterize the geometric feature of shape transitions of the trefoil knot.

\begin{figure}[h!]
        \centering
        \includegraphics[width=0.48\textwidth]{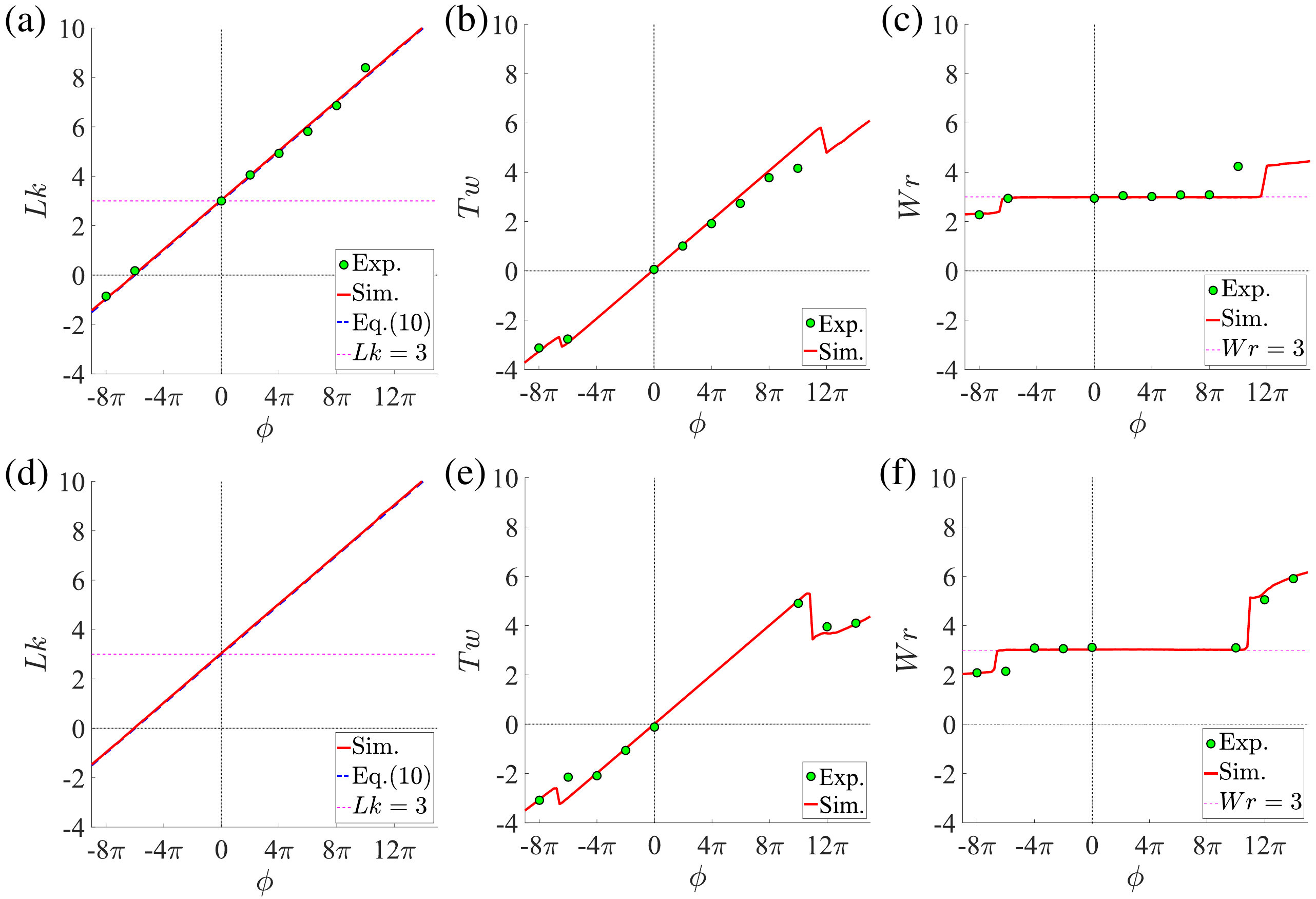}
        \caption{Topological properties for the transition of trefoil knots. The (a)(d)~twist, $Tw$, (b)(c)~writhing, $Wr$, and (c)(f)~linking, $Lk$, numbers plotted as functions of the twist angle $\phi$ for (a)-(c) $(D, L) = (8, 292)$~mm and (d)-(f) $(D, L) = (4, 280)$~mm
        The data points and red solid lines represent the experimental and simulation results, respectively. The blue dashed line in (c) and (f) is the geometrical prediction for $Lk$ as $Lk = 3+\phi/2\pi$. We plot the writhing number of the trefoil knot without twist, $Wr = 3$, as a pink dashed line for comparison.}
        \label{Fig3}
\end{figure}

Figure~\ref{Fig3}(a)--(c) presents the experimental and simulation results for the trefoil knot using a rod with $(D, L) = (8, 292)$~mm, where the horizontal axis represents the twist angle and the vertical axis represents $Lk$, $Tw$, and $Wr$, respectively. 
The experimental and numerical results of the linking number $Lk$ are in excellent agreement with our theoretical prediction Eq.~(\ref{eq:lkphi}), indicating that $Lk$ is directly related with the applied twist. In contrast, $Tw$ significantly decreases in absolute value at a certain angle, while $Wr$ increases sharply from the value of $3$ for $\phi > 0$ (or decreases for $\phi < 0$).
Notably, the transition is observed experimentally at these angles, suggesting that changes in geometric features can serve as indicators of transition occurrence.

We plot the experimental and numerical results for $(D, L) = (4, 280)$~mm in Fig.~\ref{Fig3}(d)--(f)  
The similar buckling transitions are observed at critical angles as $D = 8$~mm. In this experiment, a homogeneous elastic rod (without any inner fibers and coating surface) is used, and $Wr$ is calculated based on the same analysis as in Sec.~\ref{sec:CL} after skeletonizing the bulk rod. The twist number $Tw$ is then inferred from the geometrically determined $Lk$ using Eq.~(\ref{eq:lkphi}) and plotted accordingly. 

Although we observe the shape transitions for slender rods, the transition behaviors in the vicinity of the critical angles, $\phi^*_{\pm}$ (angles where $Tw$ or $Wr$ significantly changes), are slightly different. The agreement of $\phi^*_{\pm}$ between experiments and simulations for $D = 4$~mm is improved compared with that for $D = 8$~mm.
This discrepancy arises because the simulations consider only the centerline of the rod where the shear or cross-sectional deformations are neglected. Given that the discrepancy is reduced for slender rods, we believe that taking into account the shear or cross-sectional deformations would advance the prediction of the transition points. For example, the finite element simulation~\cite{Baek2020,Johanns2021} would be appropriate for any diameters. Nevertheless, we would investigate the properties of the transition angles using the discrete simulations based on the fact that the simulation can predict the geometric features (curvature, twist profiles, $Lk, Tw$, and $Wr$) observed in experiments.

\section{Critical twist angles for transitions}
\label{sec:angle}

\begin{figure}[h!]
        \centering
        \includegraphics[width=0.48\textwidth]{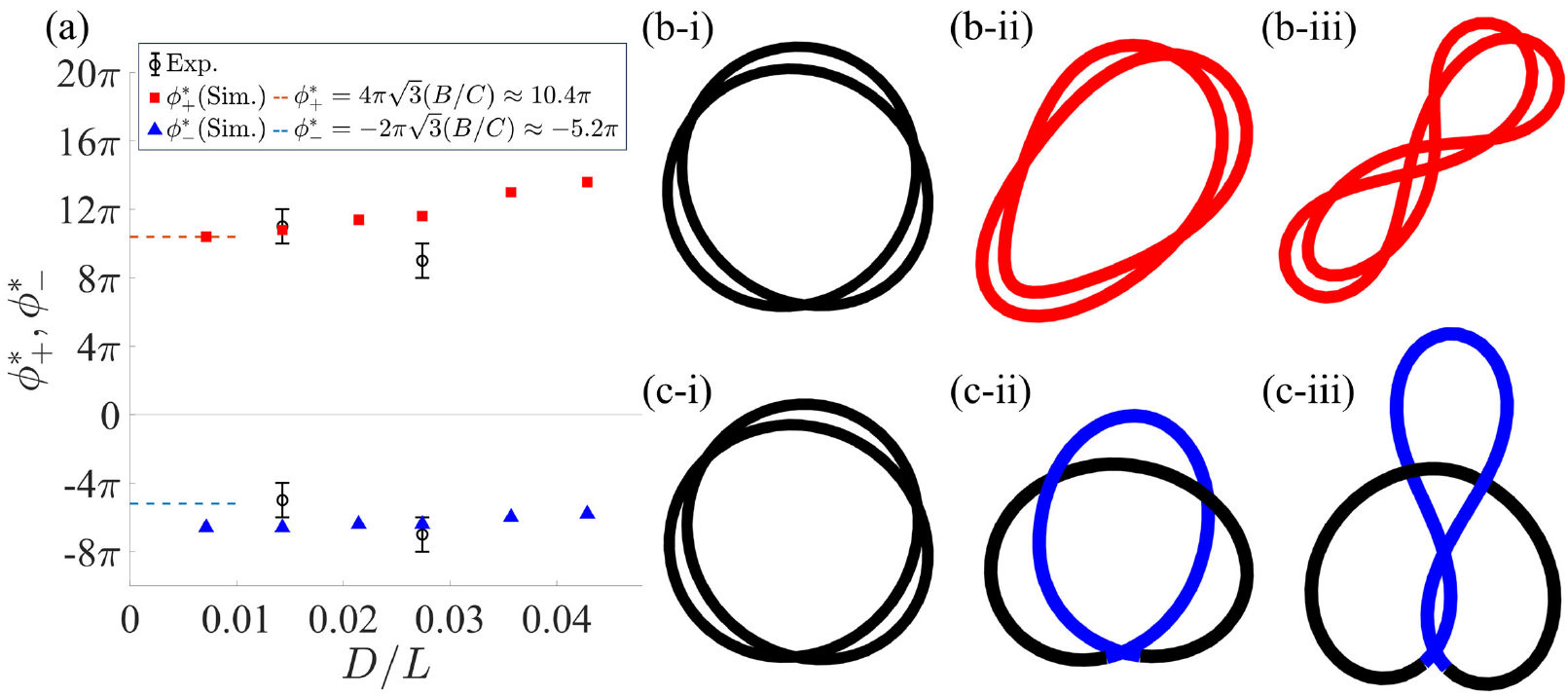}
        \caption{The instability of the twisted elastic trefoil knots. 
        (a)The critical twist angle, $\phi^*_{\pm}$, as a function of the diameter-to-length ratio, $D/L$. 
        The dashed lines represent the theoretical predictions based on equation. 
        Experimental data are plotted with the error bars of $2\pi$ (because experimental twist angle, $\phi$, has $2\pi$ precision).
        (b)(c)Snapshots of simulation with $D/L=0.0068$ in the vicinity of the transition points $\phi\simeq\phi^*_{\pm}$. 
        The applied twist angles are (b-i)$\phi=10.0\pi$, (b-ii)$\phi=10.2\pi$, (b-iii)$\phi=10.4\pi$, (c-i)$\phi=-6.2\pi$, (c-ii)$\phi=-6.4\pi$, (c-iii)$\phi=-6.6\pi$. 
        The red and blue portions would trigger each phase transition in (b) $\phi > 0$ and (c) $\phi < 0$, respectively.
        }
        \label{Fig4}
\end{figure}

By observing the changes in the geometric features of the trefoil knot calculated in Sec.\ref{sec:geometry}, we define the critical twist angle, $\phi^*_{\pm}$, as the twist angle at transition onset. For example, in the simulation results of $(D, L) = (8, 292)$~mm shown in Fig.~\ref{Fig3}(b,c) for $\phi > 0$, a decrease in $Tw$ and an increase in $Wr$ are observed. The twist angle just before this transition, $\phi = 11.6\pi$, is identified as the critical twist angle $\phi^*_+$. 
Given that we have validated the simulation against experiments, we systematically study the properties of $\phi^*_{\pm}$ for various $D/L$ (tightness) through simulation, fully relying on our numerical framework.
Fig.~\ref{Fig4}(a) presents a plot of the critical twist angles $\phi^*_{\pm}$ for both experiments and simulations across various $D/L$. We overlay the experimental results together with the error bars of $2\pi$ in Fig.~\ref{Fig4}(a).
The critical twist angle $\phi^*_+$ for $\phi > 0$ exhibits a slightly increasing trend as the diameter-to-length ratio increases. In contrast, the critical twist angle $\phi^*_-$ for $\phi < 0$ remains nearly constant, showing little dependence on the diameter-to-length ratio.

We now try to understand the transition angles defined above, $\phi^*_{\pm}$, theoretically by combining the prediction of the Michell instability (\eqref{eq:Michell}) and the topology of the knot. To clarify the detailed transition behavior, we show the simulation snapshots of a loose trefoil knot ($D/L=0.0068$) in the vicinities of transitions for $\phi > 0$ and $\phi < 0$ in Fig.~\ref{Fig4}(b) and (c), respectively. We notice that the shape of the twistless trefoil knot ($\phi = 0$, Fig.~\ref{Fig4}(b-i) and (c-i)) is regarded as that of a connected ``two rings" by neglecting the out-of-plane crossing due to the finite diameter.
When the knot is positively twisted as $\phi > 0$ (Fig.~\ref{Fig4}(b)), the entire knot (highlighted in red) folds, indicating that both overlapping rings contribute to the transition. In other words, ``two" rings buckle simultaneously into ``two" figures of eight configurations. Based on this observation, we infer that the critical twist angle at which the transition occurs for $\phi > 0$ is approximately twice the critical twist angle as
\begin{eqnarray}
    \phi^* _+ = 4\pi\kappa^{-1} \tau^* \simeq 10.4\pi,\label{eq:phip}
\end{eqnarray}
for incompressible materials $\nu = 0.5$. 
Conversely, in Fig.~\ref{Fig4}(c), only a portion of the trefoil knot undergoes buckling instability, suggesting that only ``half" of the entire structure contributes to the transition (from a single ring to a single figure-of-eight). Consequently, we predict that the critical twist angle for $\phi < 0$ would be identical to that of the Michell instability as
\begin{eqnarray}
    \phi^*_- = - 2\pi\kappa^{-1} \tau^* \simeq -5.2\pi.\label{eq:phim}
\end{eqnarray}

Although our theoretical predictions (Eqs.~(\ref{eq:phip}), (\ref{eq:phim})) are derived from simple arguments based on the Michell instability, they are found to be in good agreement with the simulation and experimental results within the limit of $D/L \ll 1$ (See dashed lines in Fig.~\ref{Fig4}(a)). This suggests that the transition angles of the trefoil knot can be predicted by considering the buckling instability of elastic rings when the knot is sufficiently loose ($D/L\ll1$).

\section{Conclusion}
\label{sec:conclusions}
\vspace{-0.3cm} 
In this study, we have investigated the effects of an applied twist on a physical trefoil knot and the resulting knot shape, revealing the mechanism of twist instability of the trefoil knot.
Our findings regarding the twist-induced buckling instabilities of the physical trefoil knot further advances the research conducted by Johann \textit{et al.}~\cite{Johanns2021}, where the shapes of the twistless trefoil knot are investigated. 
Given that the trefoil knot is one of the simple closed knots and serves the elementary structure of more complex-shaped knots, our model system study is valuable for understanding the unique mechanical properties of knots. 
The twist instability validated through our experimental analysis and discrete simulations demonstrates how the applied twist would trigger abrupt large deformation or failure of structures, thereby influencing the security of knots.
The prediction of the shape transition of the trefoil knot requires us to disentangle the complex interplay between the elasticity, geometry, and topology of the knot. Specifically, the stored twist energy of the knot is converted into bending energy upon the transition.
By introducing the geometric numbers ($Lk, Tw$ and $Wr$), we fully characterize the shape transition of the trefoil knot. We further estimate the critical transition angles, $\phi^*_{\pm}$, through the observation of the shape in the vicinity of the transition. Our phenomenological formulas for $\phi^*_{\pm}$ based on the Michell instability (buckling of the elastic ring)~\cite{michell1889, Goriely2006} are in good agreement with simulations and experiments. Still, their diameter-to-length ratio, $D/L$, dependencies need to be investigated in a more rigorous manner. We need to formulate the force and moment balance equations as well as nontrivial contact geometry and knot topology to formulate the $D/L$ dependence (e.g.,~\cite{Audoly2007, Clauvelin2009}). The $D/L$ dependencies found through our simulation would be a benchmark for future analytical models.

The mechanics of knots has been a significant research topic in the application of knot theory and in engineering applications. The increasing demand for the predictive framework of knot mechanics highlights the importance of our systematic approach, combining experiments and simulations.
The twist instability studied in this paper is predictive and holds potential for applications in various fields, including surgery, the maritime industry, and even polymer science or DNA mechanics.
We believe that controlling the twist can not only help prevent accidents and injuries but also contribute to the discovery of novel industrial designs, serving as a foundation for further engineering developments.

The possible extension of our finding is applications to the closed knots that have larger crossing numbers, $n$. The trefoil knot, the simplest prime knot, is a unique knot with a crossing number of three, $n = 3$, and is denoted as $3_1$ in the Alexander–Briggs notation~\cite{adams2004knot}. The number of prime knots of given $n$ increases with $n$ known in the knot theory~\cite{adams2004knot}. For example, there are two prime knots of $n = 5$ listed as $5_1$ (Cinquefoil knot) and $5_2$ (Three-twist knot). It would be challenging to study the twist responses of the knots with $n>3$, which we leave as intriguing future studies both experimentally and theoretically.

\newpage
\noindent \textbf{Acknowledgments.} This work was supported by MEXT KAKENHI 24H00299 (T.G.S.), JST FOREST Program, Grant Number JPMJFR212W (T.G.S.). 

\bibliographystyle{elsarticle-num-names}
\bibliography{References}

\end{document}